\documentclass{article}
\usepackage{spconf,amsmath,graphicx}
\usepackage{mathtools, amsfonts}
\usepackage{mathrsfs}
\usepackage{multicol}
\usepackage{multirow}
\usepackage{tikz}
\usepackage{tabu}
\usepackage{booktabs, tabularx}
\usepackage[referable]{threeparttablex}
\usepackage{stfloats}% for placement table on the same page

\usepackage{amsmath}
\usepackage{xcolor}
\usepackage{hyperref}
\usepackage{caption}
\usepackage{cite}
\usepackage{comment}

\usepackage{algorithm}
\usepackage{algpseudocode}

\usepackage{makecell}

\let\oldbibliography\thebibliography
\renewcommand{\thebibliography}[1]{%
  \oldbibliography{#1}%
  \setlength{\itemsep}{0mm}% use -1mm
}

\renewcommand{\vec}[1]{\boldsymbol{\mathrm{#1}}}
\newcommand{\mtx}[1]{\boldsymbol{\mathrm{#1}}}

\newcommand{\norm}[1]{\left\lVert#1\right\rVert}

\usetikzlibrary{positioning}

\newcommand*{\email}[1]{\normalsize\texttt{\href{mailto:#1}{#1}}}

\title{Probabilistic back-ends for online speaker recognition and clustering} 

\name{Alexey Sholokhov$^{1*}$\thanks{${}^{*}$Equal contribution.}, Nikita Kuzmin$^{2,3*}$, Kong Aik Lee$^3$, Eng Siong Chng$^2$ }
\address{
  $^1$Federal Research Center ``Computer Science and Control'' \\ of the Russian Academy of Sciences, Moscow, Russia \\
  $^2$Nanyang Technological University, Singapore \\
  $^3$Institute for Infocomm Research, A$^\star$STAR, Singapore \\
  \email{asholokhov@frccsc.ru}, \email{s220028@e.ntu.edu.sg},  \email{lee\_kong\_aik@i2r.a-star.edu.sg}, \email{aseschng@ntu.edu.sg}
  }
\begin{document}
\ninept % !!!!!!!!!!!!!!!!!!!!!!!!!!!!!!!!!!! <-----------------!!!!!!!!!!!!
\maketitle
%
% UNCOMMENT TO REDUCE THE TOTAL TEXT LENGTHfile:///home/alexey/Documents/work/ICASS2023_multi_enrollment/refs.bib

\setlength{\abovedisplayskip}{1pt} % 2pt <-- !
\setlength{\belowcaptionskip}{-10pt}
%%%\addtolength{\parskip}{-0.5mm}
%%% \captionsetup{belowskip=0pt}

\begin{abstract}
This paper focuses on multi-enrollment speaker recognition which naturally occurs in the task of online speaker clustering, and studies the properties of different scoring back-ends in this scenario. First, we show that popular cosine scoring suffers from poor score calibration with a varying number of enrollment utterances. Second, we propose a simple replacement for cosine scoring based on an extremely constrained version of probabilistic linear discriminant analysis (PLDA). The proposed model improves over the cosine scoring for multi-enrollment recognition while keeping the same performance in the case of one-to-one comparisons. Finally, we consider an online speaker clustering task where each step naturally involves multi-enrollment recognition. We propose an online clustering algorithm allowing us to take benefits from the PLDA model such as the ability to handle uncertainty and better score calibration. Our experiments demonstrate the effectiveness of the proposed algorithm.
\end{abstract}
%
%

% \section{Introduction}
% \label{sec:intro}
% \section{Multi-enrollment scenario}
%     \subsection{Cosine scoring with emb/score averaging}
%     \subsection{Spherical PLDA (+plda with UP?)}
%     \subsection{PSDA}
% \section{Speaker Verification}
%     \subsection{Experimental setup}
%     \subsection{Evaluation results}
% \section{Online speaker recognition (clustering?)}
%     \subsection{Experimental setup}
%         \subsubsection{Household}
%         \subsubsection{Online diarization}
%     \subsection{Evaluation results}
% \section{Qualitative Analysis}
%     \subsection{distribution of scores for multi-enroll verification}
%     \subsection{??}

%\vspace{-2mm}
\section{Introduction}
%\vspace{-1mm}
\label{sec:intro}

% Over the past several years, there has been substantial progress in developing automatic speaker recognition systems. 
% Thanks to their excessive feature extraction capabilities, deep neural networks (DNNs) have become a dominant choice in the field. DNN-based feature extractors can construct highly discriminative fixed-length vector representations, called \emph{embeddings}, from input speech signals. The similarity of these embeddings, computed via a chosen similarity measure (\emph{\emph{e.g.}}, cosine), is used to represent the similarity of speaker identities in corresponding speech utterances, which is a critical element in the decision-making process.

In this paper, we consider a general scenario that we call \emph{online speaker recognition}, where speech segments arrive sequentially, and the speaker recognition system has to identify previously encountered speakers and detect new speakers. 
At each time, there is a history of previously processed segments and the current segment to be classified.

One application scenario is \emph{household speaker recognition} \cite{Sholokhov-2022, Tan-2021}. A household is a small set of family members whose speech data is processed by a shared device such as a smart speaker (\emph{e.g.} Amazon Alexa). First, the device collects speech data from the users to create their profiles (speaker models). Then, at each interaction with a person, the device identifies the user and, optionally, updates (enriches) the corresponding speaker model. The device continuously collects the data of the users to improve its performance by estimating more accurate speaker representations. Also, the recorded speech utterances may belong to unregistered speakers (\emph{e.g.} guests) leading to an open-set identification task. Another related task is low-latency speaker spotting \cite{Patino-2018}, where a previously registered target speaker has to be detected in an audio stream.

Another example is \emph{online speaker diarization} or \emph{clustering} \cite{Liu-2004, AloniLavi-2018, Wisniewski-2017, Koshinaka-2012, Zhu-2016, Soldi-2015}. In this case, short speech segments from an audio stream have to be classified with low latency (\emph{e.g.} 1-2 seconds). 
Unlike household speaker recognition, where all unregistered speakers are not of interest, in the speaker clustering task, there are no speakers registered beforehand, and a new speaker model has to be created for each previously unseen speaker. In the following, we focus on the online speaker clustering task since it is more general, and online speaker recognition can be seen as a special case.

What these scenarios have in common is that speech segments are received \emph{sequentially} in nature and have to be classified on arrival. Specifically, an \emph{open-set identification} problem has to be solved for each new segment. That is, the current segment has to be assigned to either one of the known speakers or a new (unknown) speaker. As a result, the number of segments per speaker continuously increases over time. This requires some way to aggregate information from multiple segments to form a memory-efficient speaker representation. This is usually referred to as \emph{multi-enrollment} (or multi-session) speaker recognition \cite{Rajan2014-single-to-multiple, KALee2013-multisession, Soni-2021, Zeng-2022-att}, that is, when a speaker is represented by multiple speech segments. Moreover, different speakers may be represented by \emph{different} numbers of segments. As shown in \cite{KALee2013-multisession}, this can be a major complicating factor for speaker recognition, since it causes inconsistency in scores from different speaker models. To our best knowledge, this issue has not been studied for modern large-margin speaker embeddings.

%This problem, referred to as  partially open-set speaker detection, was studied in \cite{\cite{KALee2013-multisession}}.

% connect online speaker recognition to multi-session verification 

% Indeed, a set of previously processed speech segments is comprised of several subsets corresponding to the found speakers. The segments in a subset can be seen as the enrollment segments of a speaker. 
% The multi-enrollment verification is formulated as a hypothesis testing whether the current (yet unlabeled) segment belongs to the same speaker or not. By considering several enrolled speakers this turns into an identification problem.

% To summarize/In other words, the online speaker recognition scenario considered in this study can be seen as a sequence of multi-enrollment identification tasks. 

Inspired by \cite{KALee2013-multisession, Wang-2022}, this work focuses on the issues arising from multi-enrollment scoring since it is a core element of online speaker recognition and clustering. 
We show that popular cosine scoring could have undesirable properties when used for multi-enrollment verification. Then we show that a highly constrained version of PLDA can be a suitable alternative while having better performance and comparable computational complexity. Specifically, we propose a PLDA model with spherical between- and within-covariance matrices as a replacement for cosine scoring back-end. While being \emph{equivalent} to cosine scoring in a special case, this model can naturally handle varying degrees of uncertainty specific to the multi-enrollment scenario. 

Further, we propose a probabilistic back-end for online speaker recognition and clustering. It is based on the spherical PLDA model and therefore has several appealing properties compared to cosine scoring.
It employs an incremental (online) variant of variational Bayesian inference and provides probabilistic soft decisions for each input observation, based on the history of preceding observations.

Our contributions are summarized as follows:
\begin{itemize}
    \item We compare scoring back-ends for multi-enrollment verification for modern large-margin embeddings.
    \item We propose a simple alternative to cosine scoring suitable for multi-enrollment verification.
    \item We propose a probabilistic back-end for online speaker recognition and clustering.
    %\item uncertainty propagation
\end{itemize}

% review of [Tomi], and other papers
% There is a limited number of recent studies on this topic. 

\section{Background}
\label{sec:background}

\subsection{PLDA}

% PLDA, by-the-book scoring formula, etc
% sph-PLDA
% On equivalence to cosine
% PSDA

\textbf{General formulation.} In this study we focus on a variant of PLDA known as the \emph{two-covariance model} \cite{Brummer2010-two-cov}. Let $\vec{x}_{i,j} \in \mathbb{R}^d$ denote the $j$th speaker embedding of speaker $i$. Also, let  $\vec{y}_i$ be the latent speaker identity of speaker $i$. Then, the model is specified by two Gaussian distributions:
\begin{equation} 
    p(\vec{y}_i)  = \mathcal{N}(\vec{y}_i|\vec{\mu}, \mtx{B}), \quad p(\vec{x}_{i,j}|\vec{y}_i) = \mathcal{N}(\vec{x}_{i,j}|\vec{y}_i, \mtx{W}). \label{eq:plda}
\end{equation}
Here, $\vec{\mu}$ is a global mean, and $\mtx{B}, \mtx{W} \in \mathbb{R}^{d \times d}$ are the between- and within -speaker covariance matrices, respectively.

Being a linear Gaussian model, PLDA allows making inferences about speaker identities in closed form.
Given a set of observations (embeddings), one can compare different hypotheses about the partition of this set by computing the corresponding hypothesis likelihoods. This is often referred to as \emph{by-the-book} scoring in the literature \cite{Rajan2014-single-to-multiple, Villalba-2013-multichannel}.

\textbf{PLDA with spherical covariances.} Despite being a gold standard for previously popular i-vectors \cite{Dehak2011-ivector}, one could recently observe a gradual shift towards replacing PLDA with a simpler parameter-less cosine scoring back-end \cite{Peng-2022-plda-cos}. As discussed in \cite{Wang-2022}, the high intra-speaker compactness of the large-margin embedding makes the conventional full-rank PLDA model superfluous. It was also observed in \cite{Wang-2022} that discarding off-diagonal elements in the within-speaker covariance matrix can bring considerable performance gain. Here, we analyze a much more constrained version of the PLDA model, to our knowledge, firstly proposed in \cite{Sholokhov-2022, Kuzmin-2022}. Specifically, we consider PLDA with \emph{spherical covariances}, $\mtx{B} = \sigma_\text{B}^2\mtx{I}$, $\mtx{W} = \sigma_\text{W}^2\mtx{I}$,
% \begin{equation}
%     \begin{aligned}\label{eq:spherical-assumption}
%         \mtx{B} = \sigma_\text{B}^2\mtx{I}, \mtx{W} = \sigma_\text{W}^2\mtx{I},
%     \end{aligned}
% \end{equation}
where $\sigma_\text{B}^2$ and $\sigma_\text{W}^2$ are between- and within-speaker variances and $\mtx{I}$ denotes an identity matrix. 
%This model is parameterized by a couple of scalar parameters, which can be estimated on a given training set.
In the following text, we will refer to this model as the \emph{spherical PLDA}.

% Advantages: fast, easy to train

\textbf{Relationship with cosine scoring.} As was shown in \cite{Peng-2022-plda-cos}, for length-normalized and centered embeddings, the verification likelihood ratio of the spherical PLDA can be written as a scaled and shifted cosine similarity measure. Since an affine transformation of scores is order-preserving, the two scoring rules are equivalent. This brings up a question about the usefulness of spherical PLDA. 
As we discuss further, spherical PLDA has several advantages over cosine scoring. 
For instance, we show that the PLDA by-the-book scoring outperforms different cosine based heuristic scoring methods in multi-enrollment verification.

%For instance, in the case of multi-enrollment verification one needs to introduce
%First, it is \emph{not} equivalent to the cosine scoring in the case of multiple enrollment embeddings. To be precise, 
%one needs to introduce some heuristics since cosine similarity can be computed only for a pair of embeddings.
%unlike PLDA, cosine scoring does not naturally support 

%Can be seen as an "improved version" of the cosine scoring.

% Move to Conclusion:
% 1. Equivalent to cosine scoring (*) if we use simple embedding averaging - works as good as cosine. At the same time, cosine scoring often outperforms full-cov PLDA with modern embeddings.
% 2. Faster than diag- or full- PLDA. Easier to train - parameterized by just 2 (two) scalars.
% 3. Unlike cosine scoring, it has a natural mechanism to use information about set cardinality for multi-enrollment verification.
% 4. Unlike cosine scoring, naturally allows for uncertainty propagation (by P.Kenny) - using the input-dependent uncertainty estimates, http://www.eie.polyu.edu.hk/~mwmak/papers/UncertaintyProp.pdf
% All of the above is also true for PSDA, except that it is not exactly(!) equivalent to cosine scoring in a special case.

\textbf{Relationship with PSDA.} Another closely related scoring back-end is the so-called probabilistic spherical discriminant analysis (PSDA) recently proposed in \cite{Brummer-2022-PSDA}. It can be viewed as PLDA model with Gaussian distributions replaced by von Mises-Fisher (VMF) distributions that are defined on the $d-1$ dimensional unit hypersphere $\mathbb{S}^{d-1}$ \cite{Mardia2000-directional}:
\begin{equation} 
    p(\vec{y}_i) = \mathcal{V}(\vec{y}_i|\vec{\mu}, b), \quad p(\vec{x}_{i,j}|\vec{y}_i) = \mathcal{V}(\vec{x}_{i,j}|\vec{y}_i, w). \label{eq:psda}
\end{equation}
Here, $\mathcal{V}(\vec{y}|\vec{\mu}, \kappa)$ denotes the density of the VMF distribution with mean direction vector $\vec{\mu} \in \mathbb{S}^{d-1}$ and scalar concentration $\kappa \geq 0$ parameter. Similar to spherical PLDA, this model is parameterized by the mean direction vector $\vec{\mu}$ and two scalars: between-speaker, $b$, and within-speaker, $w$, concentrations.

%Optional paragraph
%In [Brummer] the authors show that the self-conjugacy of VMF distribution gives closed-form likelihood-ratio scores. Similar to PLDA, it is sufficient to have the average vector and set cardinality for computing posteriors over the latent variable. %They also derive an EM algorithm for estimating the parameters of this model.

% Relationship
The relation to spherical PLDA follows from the fact that restricting any isotropic Gaussian density to the unit hypersphere gives a VMF density, up to normalization. However, the two models are \emph{not} equivalent, though their behavior is very similar as we show in the experiments.  

We use both spherical PLDA and PSDA as a basis for a proposed online speaker clustering algorithm described in Section \ref{sec:main}. 

% Disadvantages of PSDA: vMF Bessel functions, 
%Despite the theoretical similarity, spherical PLDA is more attractive from the practical point of view because, unlike PSDA, it does not involve any computations with Bessel functions.

\subsection{Multi-enrollment verification}

% WHAT IS THE MAIN PURPOSE/MESSAGE OF THIS SECTION ?? Previously it was mentioned that sph-PLDA is equivalent to cosine in one-to-one case; here we should demonstrate how it is different in the many-to-one case

% Multi-enrollment verification
% Advantages of PLDA/PSDA over cosine for multi-enroll

%In this section we will discuss PLDA in the context of multi-enrollment speaker recognition. 

%We start by discussing multi-enrollment verification since it naturally occurs in online speaker clustering.
%This topic received limited attention during the recent decade.
%As the name suggests, multi-enrollment verification assumes multiple speech segments to be available 

When available, multiple enrollment utterances may represent various acoustic environments, or channels, that could be useful to better disentangle speaker identity from other irrelevant factors.

% Discuss by-the-book scoring, and why it was rarely used (inadequate enroll independence assumption, cite McCree)
% from i-vectors to modern embeddings

% https://www.iscslp2021.org/wp-content/uploads/2021/01/28.pdf
% "Multiple enrollment utterances capture a varying degree of information such as acoustic environment, channel information, language, duration, etc. and can help in improving speaker verification performance."
% "Although closed-form solution, also known as by-the-book scoring, is available for multi-enrollment scoring using PLDA models [8–10], it is rarely used for multi-enrollment speaker verification"

The study in \cite{Rajan2014-single-to-multiple} analyzes different methods of aggregating information from multiple speech segments for the PLDA scoring. Among them were embedding averaging, score averaging, and by-the-book scoring. Their experiments with i-vectors revealed that embedding averaging systematically outperforms other methods, including by-the-book scoring.
%One can show that the likelihood ratio in multi-session PLDA scoring can be rewritten as a function of the embedding average and the number of averaged embeddings (count). Therefore, the embedding averaging method is essentially the multi-session PLDA scoring with the count set to one. This corresponds to pretending that only one segment is available and the average vector represents it.
% insert some discussions from Themos' paper
%As was further discussed in \cite{McCree2017-variability}, this can be explained by an inadequate assumption of statistical independence of the enrollment segments. According to the theoretical model, enrollment segments are independent draws from the within-speaker distribution, while, in practice, it is usually not the case. %To address this problem, \cite{McCree2017-variability} proposes an extended PLDA scoring that can be seen as an intermediate case between embedding averaging and by-the-book scoring, having these two extremes as special cases. %In detail, it corresponds to setting the count parameter in the likelihood ratio score to a value between one and $N$.
%Therefore, the posterior distribution of the speaker identity variable shrinks according to:

However, these observations were made for previously popular i-vector embeddings and have not been yet confirmed for modern large-margin embeddings. In fact, in our experiments, we observe that by-the-book scoring with spherical PLDA or PSDA outperforms embedding averaging.

%Since the spherical PLDA is of primary interest in this work, we study the properties of this model in the context of multi-session scoring.

\vspace{-2mm}
\section{Online probabilistic speaker clustering}
\vspace{-1mm}
\label{sec:main}

%Structure (look at other papers):
%Online speaker clustering
%Related work
%Baseline
%Probabilistic online clustering

% MOTIVATION TO THIS SECTION 

% Difference between off-line and online clustering algorithms

In this section, we describe the proposed back-end model for online speaker recognition and clustering. The difference between offline (batch) and online settings is that in the former case all the data to be processed is available at once, while in the latter case pieces of data are observed sequentially, in some order. %In the following, we will focus only on the online clustering task.

\vspace{-2mm}
\subsection{Online clustering}
\vspace{-1mm}

%Most of the online clustering algorithms adhere to the following general pattern.
The general pattern behind many online clustering algorithms is solving a series of successive open-set identification tasks \cite{Liu-2004, Patino-2018, Wisniewski-2017, Mansfield-2018}. The basic idea is to compare each new observation to the existing clusters, and either alter the closest cluster or create a new cluster. The generic Algorithm \ref{algo:template} demonstrates this for a single time step $t$.
\vspace{-3mm}
\begin{algorithm}[H]
\caption{Online clustering (time step $t$)}
\begin{algorithmic}
%\State $\mtx{X}_1 \gets \{\vec{x}_1\}$
%\For {$t$ $\in$ $2...T$} 
\State $s_i \gets \mathrm{score}(\vec{x}_t, \mtx{X}_i)$ \Comment{Compare $\vec{x}_t$ to the existing clusters}
\State $k \gets \arg\max_i s_i$ \Comment{Find the most similar cluster}
\If{$s_k \geq \tau$} \Comment{If the maximal score $s_k$ is above the threshold $\tau$}
    \State $\mtx{X}_k \gets \{\mtx{X}_k, \vec{x}_t\}$ \Comment{Add $\vec{x}_t$ to the $k$-th cluster}
    %\State $c_k \gets \mathrm{average}(\mtx{X}_k)$ \Comment{Update the average vector}
\Else
    \State $\mtx{X}_{K+1} \gets \{\vec{x}_t\}$ \Comment{Create a new cluster}
    \State $K \gets K+1$ \Comment{Increment the total number of clusters}
\EndIf 
%\EndFor
\end{algorithmic}
\label{algo:template}
\end{algorithm}
\vspace{-5mm}
First, the observation $\vec{x}_t$ is compared to all existing clusters, each represented by a set of observations, $\mtx{X}_i$. If similarity to the closest cluster is above the threshold, $\tau$, then $\vec{x}_t$ is assigned to this cluster. Otherwise, a new cluster is formed.

% Note that each time step is essentially an open-set multi-enrollment identification task.
% Note that the algorithm involves multi-enrollment 

In this algorithm, clusters are represented by subsets of observations  sharing the same label. Therefore, computing similarity to a cluster involves many-to-one comparison, also referred to as multi-enrollment verification in the context of speaker recognition. As discussed in \cite{KALee2013-multisession}, varying cluster sizes may result in miscalibrated scores leading to sub-optimal decisions with a fixed threshold $\tau$. %In our work, we demonstrate that currently popular cosine scoring with embedding averaging also suffers from this effect.

We aim at addressing this issue and propose an algorithm suitable for online clustering. Specifically, the underlying scoring model should be robust to varying cluster sizes naturally occurring in the online scenario.
The proposed algorithm can be seen as a probabilistic extension of the Algorithm \ref{algo:template} constructed upon PLDA or PSDA models. As a result, it benefits from the advantages of PLDA (or PSDA) for multi-enrollment verification.

\subsection{Model-based clustering}

% our model

We start with a brief description of a generative model-based clustering \cite{Valente-2004, Diez2020-VBx}.

Model-based clustering builds upon a generative model that specifies how a set of data points $\mtx{X} = \{\vec{x}_1,...,\vec{x}_N\}$ is generated from the hidden parameters of $K$ clusters $\mtx{Y}=\{\vec{y}_1, ..., \vec{y}_K\}$, given the cluster assignments $\mtx{Z} = \{\vec{z}_1,...,\vec{z}_N\}$. %Here, $\vec{z}_i$ is a binary one-hot vector, whose components indicate whether the observation $\vec{x}_i$ belongs to the corresponding cluster. 
A typical clustering model is given by the following joint distribution: $p(\mtx{X}, \mtx{Y}, \mtx{Z}) = p(\mtx{X} | \mtx{Z}, \mtx{Y})p(\mtx{Y})p(\mtx{Z})$. %The cluster assignments are usually assumed to be drawn independently from the categorical distribution with prior probabilities $\pi_k$.

The clustering problem requires finding the most likely partition of the data $\mtx{Z}_* = \arg \max_{\mtx{Z}} p(\mtx{Z}|\mtx{X})$. 
%Since this task is intractable due to the huge search space, approximations are necessary. 
Our approach is based on the ``mean-field'' variational Bayesian approximation \cite{Corduneanu2001-mixture, Bishop-MachineLearning2006} assuming that the approximate posterior factorizes as $p(\mtx{Z}, \mtx{Y}|\mtx{X}) \approx q(\mtx{Z})q(\mtx{Y})$. 
%Specifically, we use the so-called ``mean-field'' approximation assuming that the approximate posterior factorizes as $q(\mtx{Z}, \mtx{Y}) = q(\mtx{Z})q(\mtx{Y})$.
This assumption leads the algorithm consisting of iterative updates of the factors $q(\mtx{Z})$ and $q(\mtx{Y})$. 
% The updates in a general form are as follows:
% \begin{equation}
%     \begin{aligned}
%         \log q(\mtx{Z}) = \mathbb{E}_{q(\mtx{Y})} [\log p(\mtx{X}, \mtx{Z}, \mtx{Y})] + \mathrm{const}, \\
%         \log q(\mtx{Y}) = \mathbb{E}_{q(\mtx{Z})} [\log p(\mtx{X}, \mtx{Z}, \mtx{Y})] + \mathrm{const},
%     \end{aligned}
% \end{equation}

% Mention VBx
%A popular speaker diarization method, sometimes referred to as VBx \cite{Diez2020-VBx}, is also an instance of the model-based clustering. 

However, such updates are designed for the conventional clustering setup, where all observations are available \emph{at once}. We modify the standard inference algorithm to make it suitable for \emph{online} clustering, where observations arrive sequentially. This algorithm can be seen as an online version of the VBx \cite{landini2020bayesian} with simplified prior on assignments $p(\mtx{Z})$. It is also similar to the algorithm from \cite{Koshinaka-2012}, where the authors modified the offline variational inference to make it suitable for online processing.
%, and predictions are made with no knowledge of future observations

%Also, it can be seen as a simplified version of the probabilistic online clustering from [], where the hidden Markov model was used
%However, if observations arrive sequentially, this algorithm should be modified

%First, we assume that, in general, predictions can be altered only within a short historical context. By default, we do not allow to alter any previous predictions.

\vspace{-2mm}
\subsection{The proposed algorithm}

Let us denote the current observation at the time step $t$ as $\vec{x}_t$, and use the notation $\mtx{X}_{1:t} = \{\vec{x}_1,...,\vec{x}_t\}$ to denote causal observations.

The algorithm updates posterior distributions of latent identity variables $q(\vec{y}_k) \approx p(\vec{y}_k|\mtx{X}_{1:t})$ after receiving a new observation $\vec{x}_t$. In general, several update iterations can be done. Our experiments reveal that even a single update can be sufficient for reasonable performance. In this case only posterior for the current data point $q(\vec{z}_t)$ needs to be computed, followed by updating each of $q(\vec{y}_k)$:
\begin{align} 
    q(\vec{z}_t) \propto \exp \sum_{k=1}^K z_{t,k} \underbrace{\left[\mathbb{E}_{q(\vec{y}_k)} [\log p(\vec{x}_t|\vec{y}_k)] + \log \pi_k\right]}_{\log \gamma_{t,k}}, \label{eq:qz}  \\
    q(\vec{y}_k) \propto \exp \left[ \gamma_{t,k} \log p(\vec{x}_t|\vec{y}_k) + \log q(\vec{y}_k|\mtx{X}_{1:t-1}) \right]. \label{eq:qy}
\end{align}
Here, $\gamma_{t,k}$ is the $k$-th component of the vector of posterior probabilities $q(\vec{z}_t)$ over the cluster assignments and $\pi_k$ are the corresponding prior probabilities.

This algorithm continuously updates speaker models defined by $q(\vec{y}_k)$. Also, one can obtain speaker labels at each time step $t$ by finding $\arg\max_k \gamma_{t,k}$. For instance, if $\gamma_{t,k}=0$, then the posterior $q(\vec{y}_k)$ stays unchanged.
%Interpretation
%If the true assignments of $\vec{x}_t$ were known, then updating $q(\vec{y}_k)$ would be nothing more than the sequential application of the Bayes formula.
%Optional paragraph:
In general, if the soft-assignments $\vec{\gamma}_t$ were converted into hard decisions, then updating $q(\vec{y}_k)$ would be nothing more than the sequential application of the Bayes formula. Also, the algorithm would become very similar to a sequence of multi-enrollment recognition tasks, where predictions are obtained via by-the-book scoring. 
% In this case predictions are computed as
% \begin{equation}
%     \begin{aligned}
%         \log q(\vec{z}_t) = \sum_{k=1}^K z_{t,k} \left[ \log \mathbb{E}_{q(\vec{y}_k)} [p(\vec{x}_t|\vec{y}_k)] + \log \pi_k \right] + \mathrm{const},
%     \end{aligned}
% \end{equation}
% where $q(\vec{z}_t) = p(\vec{z}_t | \vec{x}_t, \mtx{X}_{1:t-1}, \mtx{Z}_{1:t-1})$ is the exact posterior. One can notice the similarity to predictions obtained with variational inference (). In both cases uncertainty in estimating latent speaker identities is represented by the variance of the posterior distribution.

These update equations can be used to construct different online recognition and clustering algorithms depending on a particular choice of the underlying generative model defined by $p(\vec{x}|\vec{y})$ and $p(\vec{y})$. In this study, we use two models: spherical PLDA and PSDA. Table \ref{tab:upd} demonstrates the update equations for both models.
% $p(\vec{x}|\vec{y})p(\vec{y})$
\begin{table}[!h]
% \label{tab:upd}
\centering
\begin{tabular}{|c|c|}
\hline
\makecell{PLDA: $q(\vec{y}) = \mathcal{N}(\vec{y}|\vec{m}_t, \mtx{S}_t)$} & 
\makecell{PSDA: $q(\vec{y}) = \mathcal{V}(\vec{y}|\vec{m}_t, r_t)$} \\ \hline
\makecell{$\mtx{\Lambda}_t = \gamma_t \mtx{W}^{-1} + \mtx{\Lambda}_{t-1}$ \\ 
$\vec{\eta}_t = \gamma_t \mtx{W}^{-1}\vec{x}_t + \vec{\eta}_{t-1}$ \\
$\mtx{S}_t = \mtx{\Lambda}_t^{-1}$, $\mtx{S}_0 = \mtx{B}$ \\
$\vec{m}_t = \mtx{S}_t \vec{\eta}_t$, $\vec{m}_0 = \vec{\mu}$} & \makecell{ $\vec{\eta}_t = w \gamma_t \vec{x}_t + \vec{\eta}_{t-1}$ \\ $r_t = \norm{\vec{\eta}_t}$, $r_0=b$ \\ $\vec{m}_t = \vec{\eta}_t \mathbin{/} r_t$, $\vec{m}_0 = \mtx{\mu}$ } \\ \hline
\end{tabular}
\caption{Update equations for the full-rank PLDA \eqref{eq:plda} and PSDA \eqref{eq:psda} at the time step $t$. The speaker index is omitted for clarity.}
\label{tab:upd}
\end{table}

%About 'birth of a class' - novelty detection
To detect new speakers we introduce an extra class corresponding to an unknown speaker.
For this class, the posterior for the speaker identity variable is equal to the prior.

%Maybe include smth like Algorithm 2 from https://arxiv.org/pdf/2107.13682.pdf ?

% We demonstrate intuition behind these equations for the PLDA case, though it is similar for PSDA.
% The update equations can be re-written as follows:
% \begin{equation}
%     \begin{aligned}
%         \vec{m}_t = \alpha_t \cdot \vec{x}_t + (1 - \alpha_t) \cdot \vec{m}_{t-1}, \alpha_t \in [0, 1].
%     \end{aligned}
% \end{equation}
% Here, $\alpha_t = 0$ if $\gamma_t=0$, meaning that $\vec{m}_t$ is not updated if the current observation does not belong to the speaker $y$. Also, assuming $\gamma_i = 1$ for $i=1...t$, $\alpha_t$ behaves like the reciprocal function of $t$ that resembles incremental calculation of the average.

% FINAL ALGORITHM

%The proposed algorithm for the time step $t$ is detailed as follows.
Algorithm \ref{algo:proposed} outlines the time step $t$ of the proposed algorithms.
\vspace{-5mm}
\begin{algorithm}[H]
\caption{Proposed algorithm (time step $t$)}
\begin{algorithmic}
\State $\vec{\gamma}_t \equiv q(\vec{z}_t) \gets \text{Eq. } \eqref{eq:qz}$ \Comment{Cluster membership probabilities}
\State $q(\vec{y}_k) \gets \text{Table } \ref{tab:upd}$ \Comment{Update clusters}
\State $k \gets \arg\max_i \gamma_{t,i}$ \Comment{Find the most probable cluster}
\If{$k = K+1$} \Comment{New class is detected}
    \State $K \gets K+1$ \Comment{Increment the total number of clusters}
\EndIf 
\end{algorithmic}
\label{algo:proposed}
\end{algorithm}
\vspace{-5mm}
The advantage of the proposed algorithm over Algorithm \ref{algo:template} is that it uses soft decisions for updating clusters. This makes the algorithm more robust to classification errors.
%More importantly, as we show in our experiments, the underlying PLDA model can well handle clusters of varying sizes yielding adequate calibration of posterior probabilities, and, as a result, more accurate decisions compared to the cosine scoring. 

%It should be noted, that despite being based on PLDA, the scoring back-end model in \cite{KALee2013-multisession} was not robust to the number of enrollment utterances. This probably can be explained by the nature of i-vector distribution which is different from modern large-margin embeddings.

As a baseline for our experiments, we use Algorithm \ref{algo:template} with cosine similarity scoring.

\section{Experiments}
\label{sec:exp}

In this section, we analyze the performance of several back-end scoring models in the multi-enrollment scenario. First, we report results for a rarely investigated speaker verification scenario, \emph{i.e.}, where the number of enrollment and test segments \emph{varies} within an evaluation protocol. Next, we apply the proposed Algorithm \ref{algo:proposed} for the online speaker diarization task. To support reproducible research, we make the code and evaluation protocols publicly available.

%\subsection{Speaker embedding extractors}

% We used open-source speaker embedding extractors in order to make our experiments reproducible and computationally cheap (in terms of training time). We decided to stick to the following systems: SpeechBrain\footnote{\url{https://huggingface.co/speechbrain/spkrec-ecapa-voxceleb}} \cite{speechbrain}, BUT model\footnote{\url{https://github.com/BUTSpeechFIT/VBx}} \cite{Diez2020-VBx}, and CLOVA\footnote{\url{https://github.com/clovaai/voxceleb\_trainer}} \cite{clova}. 
% Due to space limitations we report results only for SpeechBrain, while other results can be found at github.com. 

We used open-source speaker embedding extractors in order to make our experiments reproducible. We decided to stick to the following systems: SpeechBrain \cite{speechbrain}, BUT model \cite{landini2020bayesian}, and CLOVA \cite{clova}. 
Due to space limitations, we report results only for \mbox{SpeechBrain}, while other results can be found at the project repository\footnote{\scriptsize \url{https://github.com/sholokhovalexey/online-speaker-clustering}}.

\subsection{Multi-enrollment verification}

% Goal of the experiment
% Data and protocol (our custom)
% Scoring back-ends
% Figure: distribution of scores
% Table: comparison of scoring back-ends
% Discussion

In this section, we compare different scoring back-ends in multi-enrollment speaker verification scenario.
%Specifically, we would like to 
Specifically, we investigate calibration properties of the verification scores in the case where the number of enrollment and test segment varies within an evaluation protocol.
%To our knowledge, this is the least studied scenario in speaker verification.

\textbf{Experimental setup.} We created several custom evaluation protocols from the VoxCeleb1 test set \cite{voxceleb1}. Specifically, we generated four trial lists with configurations $(1, 1)$, $(3, 1)$, $(10, 1)$, and $(3, 3)$, where the notation (\#enrollments, \#tests) represents the number of enrollment or test segments in a single trial. In addition, we combined all the trial lists to get the \emph{pooled} protocol. The idea behind it is to reveal the robustness of scoring back-ends to the number of enrollment segments. To exclude the effect of utterance duration, the recordings were cropped to 2 seconds before extracting embeddings.

%\textbf{Scoring back-ends.} 
We compared several different scoring variants: cosine similarity with embedding averaging (CSEA) or score averaging (CSSA), PSDA \cite{Brummer-2022-PSDA}, and three versions of PLDA with spherical, diagonal, and full covariance matrices. For PLDA and PSDA by-the-book scoring was used. The VoxCeleb1 dev set \cite{voxceleb1} was used for training the back-ends.
%\textbf{Metrics.} 
We used two performance metrics: the equal error rate (EER) and the minimum normalized detection cost function (minDCF) with $P_\text{target} = 0.01$ \cite{Przybocki2004-nist}. 

\begin{figure}[!h]
\centering
\includegraphics[width=8.6cm, height=2.5cm]{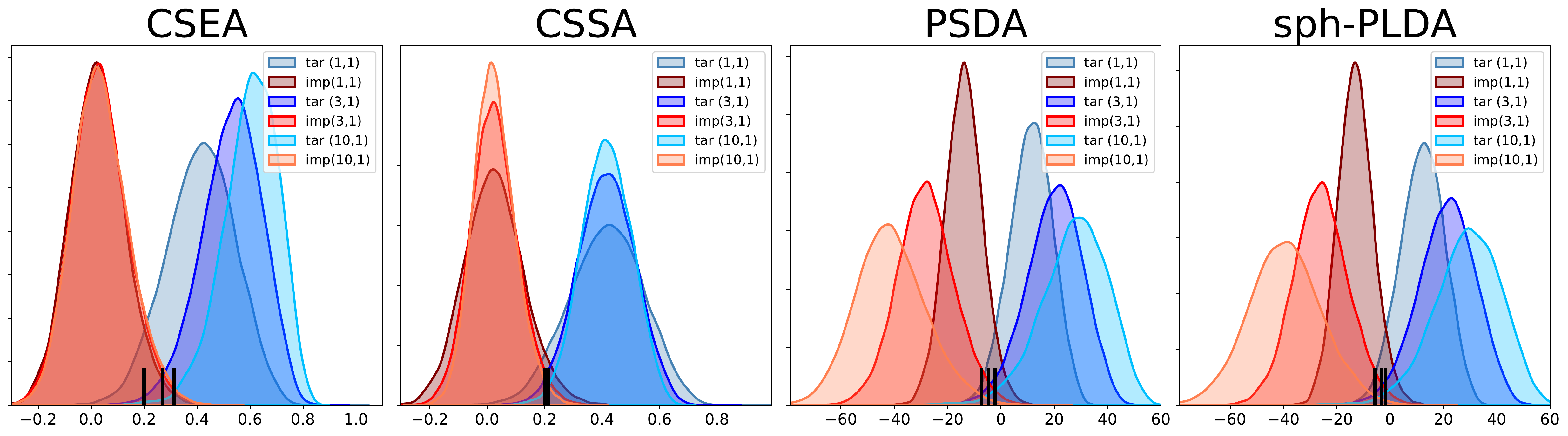}
\caption{Distributions of target and impostor scores for different numbers of enrollment segments: 1, 3, and 10. Short black vertical lines represent EER thresholds.}
\label{fig:scores}
\end{figure}

\textbf{Results.} Figure \ref{fig:scores} demonstrates the distribution of verification scores for different numbers of enrollment segments. 
%\vspace{-3mm}
One can see considerable distribution shifts for the target scores computed with CSEA. To be precise, a large variation of EER thresholds clearly makes each one sub-optimal for the other protocols.
In contrast, the EER thresholds seem to be more stable for other scoring back-ends, even despite large differences in distribution means and variances for PLDA and PSDA.
These observations are supported by objective metrics presented in Table \ref{tab:verif}.  
\begin{table}[!h]
\centering
% \begin{tabular}{|l|ccccc|}
% \hline
% \multicolumn{1}{|c|}{\multirow{2}{*}{Back-end}} & \multicolumn{5}{c|}{Protocol}                                                                                                  \\ \cline{2-6} 
% \multicolumn{1}{|c|}{}                         & \multicolumn{1}{c|}{(1, 1)} & \multicolumn{1}{c|}{(3, 1)} & \multicolumn{1}{c|}{(10, 1)} & \multicolumn{1}{c|}{(3, 3)} & pooled \\ \hline
% CSEA                                           & \multicolumn{1}{c|}{4.98}   & \multicolumn{1}{c|}{1.65}   & \multicolumn{1}{c|}{0.83}    & \multicolumn{1}{c|}{0.17}  & 2.85   \\ %\hline
% CSSA                                           & \multicolumn{1}{c|}{4.98}   & \multicolumn{1}{c|}{1.79}   & \multicolumn{1}{c|}{1.02}    & \multicolumn{1}{c|}{0.37}  & 2.05   \\ %\hline
% PSDA                                           & \multicolumn{1}{c|}{4.85}   & \multicolumn{1}{c|}{1.55}   & \multicolumn{1}{c|}{0.78}    & \multicolumn{1}{c|}{0.13}  & 2.08   \\ %\hline
% sph-PLDA                                       & \multicolumn{1}{c|}{4.98}   & \multicolumn{1}{c|}{1.59}   & \multicolumn{1}{c|}{0.78}    & \multicolumn{1}{c|}{0.14}  & 1.99   \\ %\hline
% diag-PLDA                                      & \multicolumn{1}{c|}{x}      & \multicolumn{1}{c|}{x}      & \multicolumn{1}{c|}{x}       & \multicolumn{1}{c|}{x}     & x      \\ %\hline
% full-PLDA                                      & \multicolumn{1}{c|}{x}      & \multicolumn{1}{c|}{x}      & \multicolumn{1}{c|}{x}       & \multicolumn{1}{c|}{x}     & x      \\ \hline
% \end{tabular}
\begin{tabular}{|l|ccccc|}
\hline
\multicolumn{1}{|c|}{\multirow{2}{*}{Back-end}} & \multicolumn{5}{c|}{Evaluation protocol}                                                                                                         \\ \cline{2-6} 
\multicolumn{1}{|c|}{}                         & \multicolumn{1}{c|}{(1, 1)} & \multicolumn{1}{c|}{(3, 1)} & \multicolumn{1}{c|}{(10, 1)} & \multicolumn{1}{c|}{(3, 3)} & pooled        \\ \hline
CSEA                                           & \multicolumn{1}{c|}{4.98}   & \multicolumn{1}{c|}{1.65}   & \multicolumn{1}{c|}{0.83}    & \multicolumn{1}{c|}{0.17}  & 2.85 / 0.206  \\ %\hline
CSSA                                           & \multicolumn{1}{c|}{4.98}   & \multicolumn{1}{c|}{1.79}   & \multicolumn{1}{c|}{1.02}    & \multicolumn{1}{c|}{0.37}  & 2.05 / 0.228   \\ %\hline
PSDA                                           & \multicolumn{1}{c|}{4.85}   & \multicolumn{1}{c|}{1.55}   & \multicolumn{1}{c|}{0.78}    & \multicolumn{1}{c|}{0.13}  & 2.08 / 0.172   \\ %\hline
sph-PLDA                                       & \multicolumn{1}{c|}{4.98}   & \multicolumn{1}{c|}{1.59}   & \multicolumn{1}{c|}{0.78}    & \multicolumn{1}{c|}{0.14}  & 1.99 / 0.170   \\ %\hline
diag-PLDA                                      & \multicolumn{1}{c|}{4.95}   & \multicolumn{1}{c|}{1.62}   & \multicolumn{1}{c|}{0.78}    & \multicolumn{1}{c|}{0.13}  & 1.98 / 0.169   \\ %\hline
full-PLDA                                      & \multicolumn{1}{c|}{4.74}   & \multicolumn{1}{c|}{1.79}   & \multicolumn{1}{c|}{1.08}    & \multicolumn{1}{c|}{0.20}  & 2.06 / 0.201   \\ \hline
\end{tabular}
\caption{Comparison of the speaker verification performance for different scoring back-ends in terms of EER, \%. The last column shows minDCF as well. SpeechBrain embeddings were used.}
\label{tab:verif}
\end{table}
Despite low EERs for each protocol individually, the performance of CSEA degrades significantly on the pooled protocol. In contrast, CSSA does not suffer from this problem, however, it has higher error rates on the other protocols. Finally, PLDA and PSDA perform the best, overall, handling well all the cases. They also have very similar metrics and distributions of scores. These results are also in line with findings in \cite{Sholokhov-2022} where spherical PLDA outperformed cosine similarity in the household speaker recognition task. Note that CSEA, CSSA, and sph-PLDA have exactly the same metrics in the $(1, 1)$ protocol because sph-PLDA is equivalent to cosine scoring. Another observation is that models with more parameters, diag- and full-PLDA, have comparable performance to sph-PLDA. This motivates choosing sph-PLDA as a simpler and faster alternative.

%It should be noted, that despite also being based on PLDA, the scoring back-end model in \cite{KALee2013-multisession} was not robust to the number of enrollment utterances. This probably can be explained by the nature of i-vector distribution which is different from modern large-margin embeddings.

It should be noted that, unlike this study, PLDA model studied in \cite{KALee2013-multisession} was not robust to the number of enrollment utterances. This probably can be explained by the nature of i-vector distribution which is different from the distribution of large-margin embeddings.

% +-----------+--------------+--------------+--------------+--------------+--------------+
% |  Scoring  |    (1, 1)    |    (3, 1)    |   (10, 1)    |    (3, 3)    |    pooled    |
% +-----------+--------------+--------------+--------------+--------------+--------------+
% |    CSEA   | 4.98 / 0.402 | 1.65 / 0.162 | 0.83 / 0.078 | 0.17 / 0.017 | 2.85 / 0.206 |
% |    CSSA   | 4.98 / 0.402 | 1.79 / 0.196 | 1.02 / 0.109 | 0.37 / 0.046 | 2.05 / 0.228 |
% |  sph-PLDA | 4.98 / 0.402 | 1.59 / 0.159 | 0.78 / 0.073 | 0.14 / 0.016 | 1.99 / 0.170 |
% |    PSDA   | 4.85 / 0.398 | 1.55 / 0.153 | 0.78 / 0.076 | 0.13 / 0.017 | 2.08 / 0.172 |
% | diag-PLDA | 4.95 / 0.399 | 1.62 / 0.156 | 0.78 / 0.072 | 0.13 / 0.016 | 1.98 / 0.169 |
% | full-PLDA | 4.74 / 0.459 | 1.79 / 0.187 | 1.08 / 0.107 | 0.20 / 0.022 | 2.06 / 0.201 |
% +-----------+--------------+--------------+--------------+--------------+--------------+

%\subsection{[Optional] Uncertainty propagation}

\subsection{Online speaker diarization}
% Online speaker diarization, experiments with AMI, VoxConverse

In this section, we describe experiments on online speaker diarization. We used the same PLDA and PSDA models as for the previous experiments.

\textbf{Experimental setup.} We used two popular datasets of multi-speaker recordings: AMI \cite{ami_corpus}, and VoxConverse \cite{chung20_voxconverse}. Again, due to space limitations, we report only the results for the first one, while similar observations were made for the VoxConverse.

We used the development/evaluation split for the AMI corpus from \cite{landini2020bayesian}\footnote{\scriptsize \url{https://github.com/BUTSpeechFIT/AMI-diarization-setup}}. The development set was used  for tuning the hyper-parameters of the back-end models, pretrained on the VoxCeleb data.

For AMI, the evaluation was performed on Mix-Headset channel. We extracted embeddings from segments of length $2.0$ sec with $1.0$ sec overlap within the boundaries obtained by the ground-truth annotation. These embeddings were sequentially processed by several online clustering algorithms, producing the output annotation. We did not use any special heuristics for handling segments with overlapped speakers, thus one speaker was assigned to each segment.

%We compared two versions of Algorithm \ref{algo:template}, the one with CSEA, and the other with CSSA scoring. 
We compared three versions of Algorithm \ref{algo:template}: with CSEA, CSSA, and PLDA scoring. 
Also, we evaluated two versions of the proposed Algorithm \ref{algo:proposed}, with sph-PLDA and PSDA models. All of the algorithms have at least one hyper-parameter (\emph{e.g.} decision threshold) that was tuned on the development split. 

\textbf{Results.} For the evaluation metrics, we use the diarization error rate (DER) \cite{DER_explained} and Jaccard error rate (JER) \cite{DIHARD-2}. The forgiveness collar was set to $0.25$, and overlapped speech regions were excluded from evaluation for DER, however, JER is calculated with no forgiveness collar and includes overlapped speech \cite{DIHARD-2}. %We used \texttt{pyannote.metrics} \cite{pyannote.metrics} for computing all the performance metrics.
Table \ref{tab:diar} provides the evaluation results. 
% ??? - "Note that JER considers no collar and evaluates overlap regions by definition, so it is not affected by these configurations." - ???
%TODO: run experiemnts with embeddings centering, computed on the train set (VoxCeleb)
\begin{table}[!h]
\centering
% \begin{tabular}{|l|c|c|}
% \hline
% \multicolumn{1}{|c|}{Clustering back-end} & DER, \% & JER, \% \\ \hline
% Algorithm \ref{algo:template} w/ CSEA                        & 4.95    & 9.48    \\ %\hline
% Algorithm \ref{algo:template} w/ CSSA                        & 3.68    & 10.07   \\ %\hline
% Algorithm \ref{algo:template} w/ sph-PLDA                    & 7.62    & 9.85    \\ %\hline
% Algorithm \ref{algo:proposed} w/ PSDA                        & 3.60    & 7.56    \\ %\hline
% Algorithm \ref{algo:proposed} w/ sph-PLDA                    & 3.48    & 7.69    \\ \hline
% \end{tabular}

% recalculated, skopt:
% \begin{tabular}{|l|c|c|}
% \hline
% \multicolumn{1}{|c|}{Clustering back-end} & DER, \% & JER, \% \\ \hline
% Algorithm \ref{algo:template} w/ CSEA                        & 3.86    & 25.03    \\ %\hline
% Algorithm \ref{algo:template} w/ CSSA                        & 3.66    & 26.31   \\ %\hline
% Algorithm \ref{algo:template} w/ sph-PLDA                    & 5.68    & 27.18    \\ %\hline
% Algorithm \ref{algo:proposed} w/ PSDA                        & 2.90    & 24.77    \\ %\hline
% Algorithm \ref{algo:proposed} w/ sph-PLDA                    & 2.95    & 24.00    \\ \hline
% \end{tabular}

% optuna results:

\begin{tabular}{|l|c|c|}
\hline
\multicolumn{1}{|c|}{Clustering back-end} & DER, \% & JER, \% \\ \hline
Algorithm \ref{algo:template} w/ CSEA                        & 3.63    & 25.20   \\ %\hline
Algorithm \ref{algo:template} w/ CSSA                        & 3.67    & 26.33   \\ %\hline
Algorithm \ref{algo:template} w/ sph-PLDA                    & 6.58    & 27.49   \\ \hline
Algorithm \ref{algo:proposed} w/ PSDA                        & 3.34    & 24.47   \\ %\hline
Algorithm \ref{algo:proposed} w/ sph-PLDA                    & 3.32    & 25.21   \\ \hline
\end{tabular}

% \begin{tabular}{|l|l|c|c|}
% \hline
% Mode                     & \multicolumn{1}{c|}{Clustering back-end}  & DER, \% & JER, \% \\ \hline
% \multirow{5}{*}{Online}  & Algorithm \ref{algo:template} w/ CSEA     & 4.95    & 9.48    \\
%                          & Algorithm \ref{algo:template} w/ CSSA     & 3.68    & 10.07   \\
%                          & Algorithm \ref{algo:template} w/ sph-PLDA & 7.62    & 9.85    \\
%                          & Algorithm \ref{algo:proposed} w/ PSDA     & 3.60    & 7.56    \\
%                          & Algorithm \ref{algo:proposed} w/ sph-PLDA & 3.48    & 7.69    \\ \hline
% \multirow{2}{*}{Offline} & VBx w/ sph-PLDA                           & 2.21    & 7.37    \\
%                          & VBx w/ PSDA                               & 1.99    & 10.31   \\ \hline
% \end{tabular}
\caption{Online speaker diarization with SpeechBrain embeddings.}
\label{tab:diar}
\end{table}
%Again, sph-PLDA and PSDA outperformed cosine scoring, while having similar metrics.
Unlike the previous experiment on speaker verification, we found that sequential PLDA scoring performs worse than cosine. As was discussed in \cite{McCree2017-variability} and \cite{Stafylakis2013-correlations}, this probably can be explained by an inadequate assumption of statistical independence of the enrollment segments, which affects the score calibration. According to the theoretical model, enrollment segments are independent draws from the within-speaker distribution, while in diarization it is clearly not the case because of a shared acoustic environment and recording channel. However, unlike i-vector embeddings considered in \cite{Rajan2014-single-to-multiple, McCree2017-variability}, this effect seems to be less evident for large-margin embeddings. Apparently, PLDA suffers from this effect only in diarization, while yielding adequate score calibration when embeddings are less statistically dependent, as in our previous experiment.

%PLDA model can well handle clusters of varying sizes yielding adequate score calibration, and, as a result, more accurate decisions compared to the cosine scoring. 

%It becomes evident only in diarization, while it does not affect .

%for i-vector embeddings this effect was 
%also suffers from this effect.

%To address this problem, \cite{McCree2017-variability} proposes an extended PLDA scoring that can be seen as an intermediate case between embedding averaging and by-the-book scoring, having these two extremes as special cases. 

At the same time, the proposed clustering algorithm which uses the same PLDA model delivers lower error rates than Algorithm \ref{algo:template}. In the future, we plan to further investigate the properties of this algorithm in other applications such as household speaker recognition, where speech utterances are also processed sequentially.
%These results suggest that the prediction rule \label{eq:qz} is more suitable for online recognition where performance depends on the \emph{order} of arriving observations.
%optional: offline
%For reference, we also implemented two versions of the \emph{offline} VBx model \cite{landini2020bayesian} based on sph-PLDA and PSDA. The results are comparable to those reported in the original study \cite{landini2020bayesian} but for the other embeddings.

%We should note, that this algorithm is general and can be used with any large-margin embeddings.

\section{Conclusion}
%This paper lies at the intersection of two topics: multi-enrollment speaker verification and scoring methods for modern large-margin speaker embeddings.
%This paper explores the multi-faceted/composite topic that combines multi-enrollment speaker verification and scoring methods for modern large-margin speaker embeddings.
This paper studies the properties of popular scoring back-ends suitable for large-margin speaker embeddings, with a particular focus on multi-enrollment speaker verification. Our experiments with the state-of-the-art embeddings revealed shortcomings of cosine scoring in the multi-enrollment scenario. To address this, we advocate for using the spherical PLDA that has several attractive properties: absence of numerical instabilities specific to PSDA due to Bessel functions; better performance, comparable computational complexity, and equivalence to cosine scoring in a special case. 
%Also, viewing online clustering as a sequence of open-set speaker identification tasks, we introduced a simple algorithm that uses the advantages of PLDA and PSDA for the multi-enrollment scenario. 
Also, we introduced a simple online clustering algorithm that uses the advantages of PLDA and PSDA for the multi-enrollment scenario. 
Empirical evaluation of the online speaker diarization showed superior performance of the proposed algorithm.


\vfill\pagebreak


% https://tex.stackexchange.com/questions/131087/displaying-authors-name-in-a-bibliographic-entry-in-the-form-surname-first-in
% line 231

\bibliographystyle{IEEEbib}
\bibliography{refs}

\end{document}